\documentclass[aip,jcp,reprint,amsmath,amssymb]{revtex4-2}
\usepackage{graphicx}

\newcommand{\kF}{k_{\mathrm F}}
\newcommand{\TF}{T_{\mathrm F}}
\newcommand{\de}{\Delta\varepsilon}
\newcommand{\Sx}{\Sigma_x}
\newcommand{\bk}{\mathbf{k}}
\newcommand{\bq}{\mathbf{q}}

\begin{document}

\title{For any static interaction the self-energy and vertex corrections to the electron-gas polarizability combine exactly into one integrand free of the Coulomb singularity}

\author{Carl A. Kukkonen}
\email{kukkonen@cox.net}
\affiliation{33841 Mercator Isle, Dana Point, California 92629, USA}

\begin{abstract}
The first-order corrections to the polarizability of the uniform electron gas are a self-energy insertion and a vertex correction. For the Coulomb interaction the coefficient of $q^2$ in each diverges logarithmically at small wave vector, and the two divergences have opposite signs and cancel in their sum. This Note shows that for any interaction that depends only on the momentum it transfers, the two integrands combine exactly into one before any integration is done. At zero temperature the combined integrand has no Coulomb singularity and no chemical-potential counterterm. For screened interactions, which must be integrated numerically, it should be computed in place of the two diagrams. The Note gives the derivation and its physical reason, the one term added at finite temperature, and checks for bare, Yukawa and screened interactions. With a frequency-dependent interaction the combination is no longer exact.
\end{abstract}

\maketitle

The first-order corrections to the random phase approximation polarizability of the electron gas are three diagrams of two kinds, each with one interaction $W$, Figs.~1(a) to 1(c). In the self-energy insertion one state of the excited electron-hole pair exchanges once with the Fermi sea, the electron in (a) or the hole in (b). In the vertex correction (c) the electron and the hole interact once with each other. For the interactions used here the two kinds have opposite signs, because exchange raises the energy cost of every excitation and the electron-hole attraction lowers it. For the bare Coulomb interaction the coefficient of $q^2$ in each diverges as $\ln q$ at small $q$, while their sum is the exact exchange-only local field factor $G_x$,\cite{Kukkonen_JCP} whose coefficient is one quarter, the exchange correction to the compressibility. The same near-cancellation has been seen in semiconductors\cite{Ummels} and in molecules, where it was called ``still poorly understood'' in 2024.\cite{Forster}

The proper polarizability, the response with the Coulomb chain of the RPA removed, is $\chi=\chi_0+\chi_1+\dots$, with $\chi_0$ the free response and $\chi_1$ the correction of first order in $W$. It defines the local field factor $G$, the correction to the RPA, through $G=-\chi_1/(V_C\chi_0^2)$, with $V_C(q)=4\pi e^2/q^2$. Densities are given by $r_s$, the mean electron spacing in Bohr radii, and temperatures by $T/\TF$ with $\TF$ the Fermi temperature. A probe of wave vector $\bq$ and zero frequency mixes into each occupied state $\bk$ a small amount of the state $\bk+\bq$, in proportion to $1/\de_k$, where $\de_k=\varepsilon_{k+q}-\varepsilon_k$ is the energy cost of the move and $\varepsilon_k=\hbar^2k^2/2m$. Write $f_k=n_k-n_{k+q}$ for the occupancy difference of the pair, so that $\chi_0=-2\int_k f_k/\de_k$, with $\int_k\equiv\int d^3k/(2\pi)^3$ and the factor 2 for spin. For an interaction $W(|\bk-\bk'|)$ that depends only on the momentum it transfers, the two diagrams and their sum are, exactly,
\begin{subequations}
\begin{align}
\chi_{\mathrm{SE}}(q)&=\int_k\!\int_{k'}W(\bk-\bk')\,f_kf_{k'}\left[\frac{1}{\de_k^2}+\frac{1}{\de_{k'}^2}\right],\label{eq:SE}\\
\chi_{\mathrm{V}}(q)&=-2\int_k\!\int_{k'}W(\bk-\bk')\,\frac{f_kf_{k'}}{\de_k\,\de_{k'}},\label{eq:V}\\
\chi_1(q)&=\int_k\!\int_{k'}W(\bk-\bk')\,f_kf_{k'}\left[\frac{1}{\de_k}-\frac{1}{\de_{k'}}\right]^2.\label{eq:sq}
\end{align}
\end{subequations}

The three differ only in the bracket. Two electrons with the same momentum have the same energy cost, so the bracket of Eq.~(\ref{eq:sq}) vanishes exactly where $W$ is largest. For the bare interaction $W$ is infinite there, and that singularity makes each diagram diverge as $\ln q$. In Eq.~(\ref{eq:sq}) the product stays bounded and the two divergences never form.

Equation~(\ref{eq:SE}) does not look like a self-energy diagram, and putting it in that form is the whole derivation. Exchange shifts the energy cost of the pair by $\Sx(\bk+\bq)-\Sx(\bk)$, the shift of the electron's level minus the shift of the level it left, where the hole is. With $\Sx(\bk)=-\int_{k'}W(\bk-\bk')\,n_{k'}$, write that difference out and rename $\bk'\to\bk'+\bq$ in the first of its two integrals. The interaction becomes $W(\bk-\bk')$ in both, the two occupations combine, and
\begin{equation}
\Sx(\bk+\bq)-\Sx(\bk)=\int_{k'}W(\bk-\bk')\,f_{k'}.\label{eq:rename}
\end{equation}
The self-energy diagram is $2\int_k f_k[\Sx(\bk+\bq)-\Sx(\bk)]/\de_k^2$, and with Eq.~(\ref{eq:rename}), made symmetric in $\bk$ and $\bk'$, it becomes Eq.~(\ref{eq:SE}). The step uses only two facts. Shifting both momenta by $\bq$ leaves $W(\bk-\bk')$ unchanged, and the integral over $\bk'$ is unrestricted. Nothing refers to the $1/k^2$ of the Coulomb interaction, so the result holds for bare, Yukawa and screened interactions alike. The interaction sees only relative momentum, so a common shift of the whole Fermi sea costs no exchange energy and only an electron's motion relative to the others does. A constant added to $\Sx$ drops out of Eq.~(\ref{eq:rename}), so the chemical-potential counterterm $\delta\mu$, which a calculation at fixed density subtracts from the self-energy, never appears in Eq.~(\ref{eq:sq}).

Parts of this identity are in print. Geldart and Taylor's Eq.~[24] is Eq.~(\ref{eq:sq}) for the bare Coulomb interaction, written to make a numerical integration converge.\cite{GT} The small-$q$ cancellation,\cite{GV} the two diagrams in closed form with the cancellation credited to DuBois,\cite{EV,DuBois} and Eqs.~(\ref{eq:rename}) and (\ref{eq:sq}) for the bare two-dimensional gas\cite{DasSarma} are also in print. What this Note adds is the extension to any static interaction, stated and checked, the one term at finite temperature, and the recommendation for screened interactions.

At finite temperature one term is added. Exchange shifts each level by $\Sx(\bk)-\delta\mu$, with $\delta\mu$ the shift of the chemical potential that holds the density fixed, $\int_k n'_k[\Sx(\bk)-\delta\mu]=0$. At zero temperature $n'$ is a delta function on the Fermi surface. In the uniform gas $\Sx$ is constant there and equals $\delta\mu$, so the term vanishes and no electron crosses. At finite temperature the step is smeared over a width of order $T$, $\Sx$ varies across that width, and the self-energy pushes some electrons across the Fermi surface. The occupation of state $\bk$ changes by $n'_k[\Sx(\bk)-\delta\mu]$, with $n'$ the derivative of the Fermi function. The free gas responds to that change, and with Fermi functions in $f_k$ and $f_{k'}$
\begin{multline}
\chi_1(q,T)=\text{Eq.~(\ref{eq:sq})}\\
-2\int_k\frac{n'_k[\Sx(\bk)-\delta\mu]-n'_{k+q}[\Sx(\bk+\bq)-\delta\mu]}{\de_k}.\label{eq:T}
\end{multline}

The vertex diagram is built on unperturbed occupations and has nothing to pair with this term. For a static interaction in the uniform gas it is the only departure from Eq.~(\ref{eq:sq}), and $\delta\mu$ enters only through it.

\begin{figure}
\includegraphics[width=0.84\columnwidth]{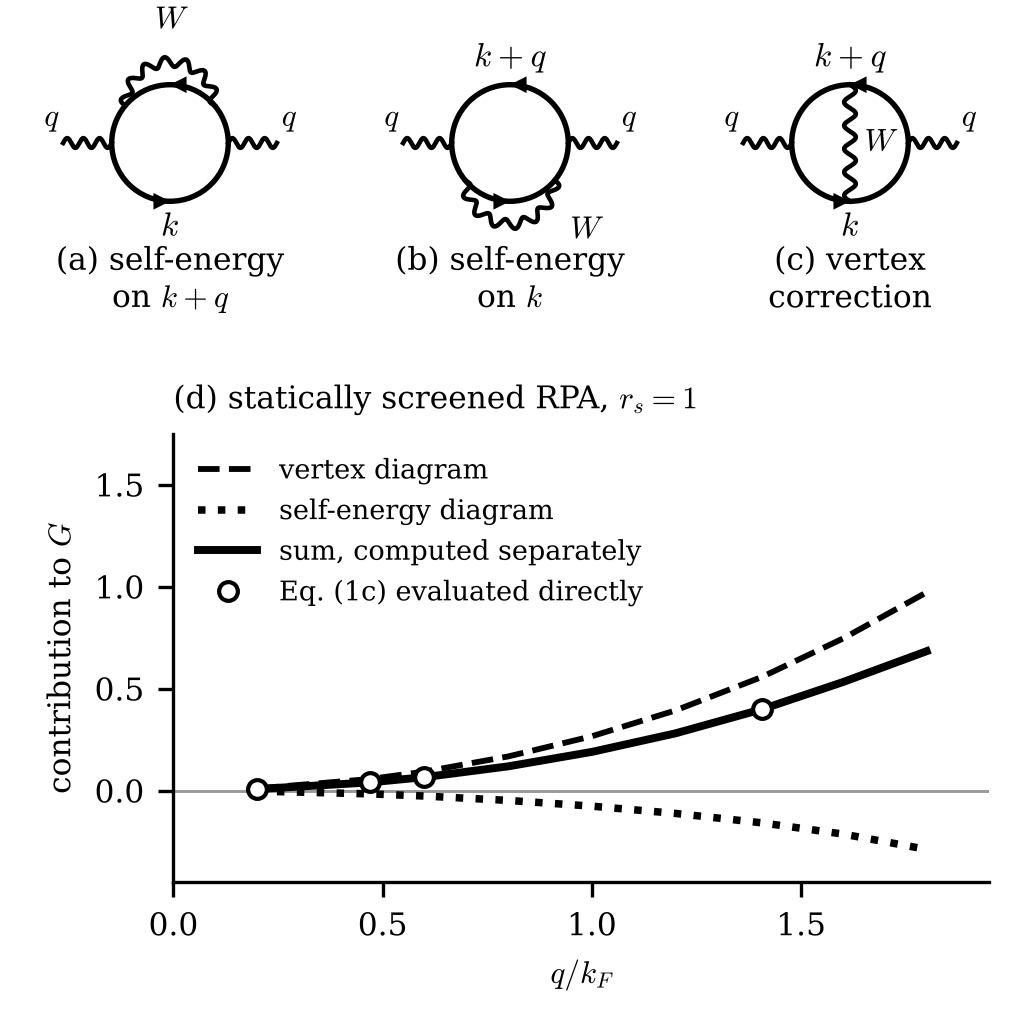}
\caption{(a) to (c) The two first-order diagrams. The outer wavy lines are the probe and the inner one is $W$. The self-energy is inserted on the upper state $\bk+\bq$ (a) or the lower state $\bk$ (b), and both are in Eq.~(\ref{eq:SE}). (c) The vertex correction. (d) The statically screened RPA interaction at $r_s=1$. Curves, the two diagrams and their sum computed separately at $T/\TF=0.025$. Open circles, Eq.~(\ref{eq:sq}) at $T=0$ where it was evaluated. The curves stop below the cusp at $2\kF$.}
\end{figure}

The supplementary material checks the combined form against references that share no code with it. At zero temperature Eq.~(\ref{eq:sq}) agrees with $G_x$ in closed form for the bare interaction to 0.013 percent or better, and with the two diagrams computed separately for Yukawa interactions to 0.001 percent. At $T/\TF=0.025$ the self-energy diagram computed with Eq.~(\ref{eq:T}) agrees to 0.003 percent with the diagram in its original form, which carries the self-energy and $\delta\mu$ on the propagators. For the statically screened RPA interaction the reference is at $T/\TF=0.025$ and the agreement is at the 0.1 percent level, Fig.~1(d).

In the author's implementation Eq.~(\ref{eq:sq}) reached a given accuracy 38 to 66 times faster than the two diagrams computed separately on a product grid without subtraction of the Coulomb singularity, and reached accuracies the latter did not. One caution applies. The integrand of Eq.~(\ref{eq:sq}) still varies steeply near the circle where $\bk$ and $\bk+\bq$ both lie on the Fermi surface, where the free response has its own integrable $1/\de_k$, and the grid must crowd onto that circle. With that, for any static interaction, compute Eq.~(\ref{eq:sq}), not the two diagrams.

For an interaction that depends on frequency, such as the dynamically screened RPA interaction, the combination is no longer exact, because the renaming of Eq.~(\ref{eq:rename}) needs $W$ to depend only on momentum, but the frequency dependence is essential for agreement with quantum Monte Carlo results for the local field factor, and most of the cancellation survives it.\cite{Letter} That case will be treated separately.

\medskip
\noindent\textbf{ACKNOWLEDGMENTS.} The author conceived the question, directed the work and reviewed every step. The derivation, code, calculations and drafts were produced with Claude, an AI system from Anthropic (Opus 5 and Fable 5 models, October 2026). The author is solely responsible for the content.

\noindent\textbf{SUPPLEMENTARY MATERIAL.} See the supplementary material for the table of checks.

\noindent\textbf{AUTHOR DECLARATIONS.} \textit{Conflict of Interest.} The author has no conflicts to disclose.

\noindent\textbf{DATA AVAILABILITY.} The data and code that support this study are available from the author upon reasonable request.

\clearpage
\onecolumngrid
\section*{Supplementary material}
Table S1 checks the combined form, Eq.~(1c) of the Note, against references that share no code with it. The reference is named in each block. The finite-temperature reference computes the self-energy diagram in its original form, with the self-energy and $\delta\mu$ on the propagators, and never uses Eq.~(3) of the Note. At equal temperatures the two sides agree to between 0.001 and 0.013 percent. For the statically screened RPA interaction Eq.~(1c) was evaluated at $T=0$ and the reference at $T/\TF=0.025$, and the agreement is at the 0.1 percent level.

\begin{table}[h]
\renewcommand{\thetable}{S1}
\caption{The combined form against independent references, as contributions to $G$. In (a) to (c) the quantity is the sum of the two diagrams, Eq.~(1c) of the Note. In (d) it is the self-energy diagram alone, computed as its difference form plus the added term of Eq.~(3) of the Note, which is where that term enters.}
\label{tab:checks}
\begin{ruledtabular}
\begin{tabular}{lcccc}
interaction & $q/\kF$ & this Note & reference & diff., \% \\
\hline
\multicolumn{5}{p{0.85\textwidth}}{(a) Bare Coulomb, $T=0$. Reference, $G_x$ in closed form.}\\
bare & 0.4 & 0.040912 & 0.040910 & $+0.004$ \\
bare & 1.4 & 0.67743 & 0.67734 & $+0.013$ \\
\multicolumn{5}{p{0.85\textwidth}}{(b) Yukawa, $W=4\pi e^2/(k^2+\kappa^2)$, $T=0$. Reference, the two diagrams separately. The same holds at $\kappa=2\kF$.}\\
$\kappa=0.5\kF$ & 1.0 & 0.224305 & 0.224308 & $-0.001$ \\
\multicolumn{5}{p{0.85\textwidth}}{(c) Statically screened RPA. Eq.~(1c) of the Note at $T=0$, reference the two diagrams at $T/\TF=0.025$.}\\
$r_s=1$ & 1.0 & 0.189253 & 0.189114 & $+0.073$ \\
$r_s=5$ & 1.0 & 0.102240 & 0.102190 & $+0.049$ \\
\multicolumn{5}{p{0.85\textwidth}}{(d) Bare Coulomb, $T/\TF=0.025$, self-energy diagram with the term of Eq.~(3) of the Note. Reference, the diagram in its original form.}\\
bare & 0.4 & $-0.1156916$ & $-0.1156956$ & $-0.003$ \\
bare & 1.4 & $-1.1485736$ & $-1.1486002$ & $-0.002$ \\
\end{tabular}
\end{ruledtabular}
\end{table}
\end{document}